\documentclass[conf]{new-aiaa}
\usepackage[utf8]{inputenc}

\usepackage{graphicx}
\usepackage{amsmath}
\usepackage[version=4]{mhchem}
\usepackage{siunitx}
\usepackage{longtable,tabularx}
\DeclareMathOperator*{\argmax}{arg\,max}

\title{When to compute in space}

\author{Rajiv Thummala \footnote{PhD Student,  Sibley School of Mech. and Aerospace Engineering, 124 Hoy Rd, and AIAA Member} and Gregory Falco \footnote{Assistant Professor, Department of Mech. and Aerospace Engineering, 124 Hoy Rd, and AIAA Member}}
\affil{Cornell University, Ithaca, NY, 14850}

\begin{document}

\maketitle

\begin{abstract}
Spacecraft increasingly rely on heterogeneous computing resources spanning onboard flight computers, orbital data centers, ground station edge nodes, and terrestrial cloud infrastructure. Selecting where a workload should execute is a nontrivial multi objective problem driven by latency, reliability, power, communication constraints, cost, and regulatory feasibility. This paper introduces a quantitative optimization framework that formalizes compute‐location selection through empirically measurable metrics, normalized scoring, feasibility constraints, and a unified utility function designed to operate under incomplete information. We evaluate the model on two representative workloads demonstrating how the framework compares compute tiers and identifies preferred deployment locations. The approach provides a structured, extensible method for mission designers to reason about compute placement in emerging space architectures.
\end{abstract}

\section{Introduction}
Modern satellites generate unprecedented volumes of data. Historically, this data followed the simple flow of being captured through the payload, stored in onboard memory, and downlinked to ground stations for processing. This paradigm is shifting, however, as new computational paradigms and space infrastructure is enabling diverse options for where computation occurs. In the decades to come, mission architects can choose to perform computation for a spacecraft on the onboard flight computer, an orbital data center, ground station edge nodes, and terrestrial data centers. However, the decision of where to execute a computational workload in the context of space systems remains a complex multi-objective optimization problem. Each compute tier presents distinct trade-offs across latency, reliability, power consumption, communication overhead, and cost. Furthermore, regulatory considerations such as ITAR restrictions, data sovereignty requirements, and mission security policies may impose hard constraints that supersede performance considerations.

Existing approaches to compute placement in distributed systems typically assume homogeneous network conditions and focus on cloud-edge trade-offs in terrestrial settings. These frameworks do not adequately capture the unique constraints of space systems such as intermittent line-of-sight communication windows, orbit-dependent propagation delays, power budgets dominated by solar panel geometry and eclipse periods, radiation effects on computational reliability, and the need to balance operational complexity against mission autonomy.

This paper \cite{ThummalaFalco2026WhenToComputeInSpace} introduces a formal optimization framework designed specifically for this heterogeneous space compute landscape, providing mission architects with a principled method to navigate these trade-offs and select the appropriate compute tier for each workload.

\section{Prior Art}
Several studies describe the ongoing shift in how computing is conceived and implemented in the space domain. \cite{liu2025computing} outlines four principal phases in the development of space computing architectures. The first, distributed embedded systems, characterizes spacecraft composed of independent embedded units. The second, integrated electronic systems, consolidates multiple functions onto a single platform. The third, external intelligent systems, augments the integrated architecture with high performance computing modules capable of executing complex algorithms, including artificial intelligence. The fourth, integrated intelligent systems, unifies the external and integrated architectures into a single construct with reduced power consumption, smaller volume, and improved performance. Although the work provides a clear articulation of the advantages and limitations of each paradigm, it does not supply a formal decision-making construct that would allow mission engineers to determine which architecture is most appropriate for a given application.

\cite{peraspera2025realities} similarly compares multiple space computing paradigms and examines their operational constraints, incorporating examples from real systems such as the James Webb Space Telescope. The authors argue that an orbital data center located at the Earth–Moon L1 point could support activities such as calibration, cosmic ray scrubbing, and first-order science processing by receiving raw instrument data directly from JWST. This paper, however, also stops short of offering a rigorous decision function to guide engineers in selecting an optimal computing architecture.

Although academic literature on orbital data centers remains sparse, the concept of employing sun-synchronous orbits to provide large-scale AI computation and inference has gained widespread media attention. \cite{callison2025beyond} highlights that orbital solar power emerges as a cost-saving cornerstone for data centers, addressing the staggering energy burdens of terrestrial facilities where electricity devours 40\% to 60\% of annual expenses—up to \$5-10 million yearly for hyperscale operations. In sun-synchronous orbits, solar panels thrive under uninterrupted sunlight, unhindered by clouds, pollution, or atmospheric filtering, basking in an irradiance of 1,366 W/m² compared to Earth's diluted average. For instance, deploying 1,000 kg of arrays via Falcon 9 at \$2,500 per kg incurs a mere \$2.5 million upfront, yielding 500 kW of capacity at roughly \$5,000 per kW, which is on par with ground-based setups when amortized over extended lifecycles. This shift slashes annual operational expenditures to zero post-deployment, swaps water-guzzling cooling (4-5 million gallons daily) for efficient radiative methods, and eases regulatory hurdles from zoning and land grabs to streamlined ITU spectrum and orbital licensing. Ultimately, solar-powered orbital platforms not only eradicate millions in energy and cooling costs but also propel sustainability goals, redefining data center economics for a 10- to 20-year horizon.

Several organizations have proposed architectures based on this model. Google Suncatcher \cite{fischbacher2025towards} seeks to deploy solar-powered satellites carrying TPU accelerators, with the number and size of TPUs determined by engineering and economic constraints. Their design places satellites in a dawn–dusk sun-synchronous low Earth orbit to maximize solar power availability while keeping ground communication latency and launch costs low. Their stated rationale rests on the scale of solar output, which exceeds global human electricity consumption by more than fourteen orders of magnitude, implying that future large-scale AI workloads may be most efficiently powered by direct access to solar flux. Other entities, including Starcloud \cite{feilden2024why} and Axiom Space \cite{axiomspace2025inspace}, are developing orbital data center concepts intended for AI training workloads. Axiom Space asserts that such an architecture provides vastly greater volume, external mounting area, electrical power, and heat rejection capacity, along with a benign pressurized environment that supports commercial-grade hardware. They further argue that routine crewed and resupply missions would enable economical delivery of upgrades, maintenance, and physical commodities to LEO \cite{axiomspace2025inspace}.

Taken together, these works constitute a small fraction of the emerging discourse on space-based computation, whether academic or media-driven. They nonetheless lack formal structure and do not identify the empirical metrics needed to determine when an orbital data center, rather than terrestrial infrastructure, represents the optimal solution for a given mission profile.

\section{Compute-Location Optimization Model}
\label{sec:compute_opt_model}


The ability for a spacecraft to offload compute to heterogeneous resources has received significant attention in recent literature. This can include the onboard computer, orbital data centers, ground station edge nodes, and terrestrial cloud infrastructure. Determining where a process should execute involves balancing competing objectives including latency, reliability, power consumption, communication overhead, cost, and regulatory feasibility. To formalize this decision process, we introduce a constrained multi criteria optimization model that evaluates various computational tiers using a unified utility function. The selected compute location is the one that maximizes the effective utility subject to the spacecraft's mission constraints.

\subsection{Variables and Notation}

Let \( j \in \{\mathrm{FC}, \mathrm{ODC}, \mathrm{GSE}, \mathrm{TDC}\} \) denote a compute tier: onboard flight computer (FC), orbital data center (ODC), ground station edge (GSE), or terrestrial data center (TDC). Each tier is described by a vector of empirically measurable metrics. Let \( i \in I \) index the set of all criteria. Table~\ref{tab:variables} summarizes the variables used in the optimization formulation.

\begin{table}[h!]
\centering
\caption{Metrics and variables used in the compute-location optimization model.}
\label{tab:variables}
\begin{tabular}{ll}
\hline
\textbf{Symbol} & \textbf{Description} \\ \hline
\( M_i(j) \)                & Raw value of metric \( i \) for tier \( j \) \\
\( w_i \)                   & Weight of metric \( i \) in the utility computation \\
\( s_i(j) \)                & Normalized score of metric \( i \) for tier \( j \), in \([0,1]\) \\
\( m_i^{\min}, m_i^{\max} \)& Expected operating bounds of metric \( i \) \\
\( K_j \subseteq I \)       & Set of metrics available for tier \( j \) (some may be missing) \\
\( W_{\text{total}} \)      & Total weight sum, \( W_{\text{total}} = \sum_{i \in I} w_i \) \\
\( W_{\text{known}}(j) \)   & Weight sum of metrics available for tier \( j \) \\
\( \phi(j) \)               & Information fraction for tier \( j \), \( \phi(j) = W_{\text{known}}(j)/W_{\text{total}} \) \\
\( U_{\text{base}}(j) \)    & Base utility computed from known metrics \\
\( U_{\text{eff}}(j) \)     & Penalized utility after missing-data adjustment \\
\( D_{\text{req}} \)        & Task requirement: maximum allowable latency \\
\( R_{\text{req}} \)        & Minimum acceptable success probability \\
\( Q_{\text{req}} \)        & Minimum acceptable output quality \\
\( C_{\text{req}} \)        & Maximum allowable marginal cost per task \\
\( R_{\text{ok}}(j) \)      & Regulatory feasibility indicator \((0~\text{or}~1)\) \\
\hline
\end{tabular}
\end{table}

The metric set includes the following empirically measurable quantities:

\begin{itemize}
\item End-to-end p99 latency \( M_{\text{latency}}(j) \)
\item Success probability \( M_{\text{success}}(j) \)
\item Output quality \( M_{\text{quality}}(j) \)
\item Energy per task \( M_{\text{energy}}(j) \)
\item Peak power draw \( M_{\text{peak\_power}}(j) \)
\item Power margin \( M_{\text{power\_margin}}(j) \)
\item Link availability \( M_{\text{link}}(j) \)
\item Contact duty cycle \( M_{\text{contact}}(j) \)
\item Data reduction ratio \( M_{\text{reduction}}(j) \)
\item Marginal cost per task \( M_{\text{cost}}(j) \)
\item Operational overhead \( M_{\text{ops}}(j) \)
\item Compute availability \( M_{\text{comp\_avail}}(j) \)
\item Orbital altitude \( M_{\text{alt}}(j) \)
\item Compute subsystem mass \( M_{\text{mass}}(j) \)
\item Regulatory feasibility \( R_{\text{ok}}(j) \)
\end{itemize}

All metrics are scalar and can be measured experimentally or obtained from system models.

\subsection{Rationale for Optimization Variables}
\label{sec:metric_rationale}

The metrics included in the optimization model capture the fundamental physical, operational, and economic factors that determine whether a computation should occur onboard, in orbit, at the ground station edge, or in terrestrial cloud infrastructure. Each variable represents a quantitatively measurable property that affects system performance or mission feasibility. The following subsections describe the rationale for each metric category and its relevance to compute-location selection.

\subsubsection{Latency and Timing Metrics}

\textbf{End-to-end p99 latency} \( M_{\text{latency}}(j) \)  
captures the worst-case delay experienced during time-critical operations. Many spacecraft functions such as fault detection, autonomy, and onboard intrusion detection require deterministic responses within strict deadlines. Lower latency strongly favors onboard computation, while higher values may render remote computation infeasible.

\textbf{Median latency} \( M_{\text{latency50}}(j) \)  
characterizes typical responsiveness and is useful in tasks with soft-real-time requirements. Although less restrictive than p99 latency, it provides insight into nominal compute performance.

\textbf{Latency jitter} (optional) captures the variability between typical and worst-case timing conditions. High jitter reduces predictability and may disqualify a tier from hosting tightly coupled control functions.

\subsubsection{Reliability, Correctness, and Safety Metrics}

\textbf{Success probability} \( M_{\text{success}}(j) \)  
is the likelihood that a computation completes both correctly and within its deadline. This metric is essential for safety-critical tasks where probabilistic guarantees must be satisfied.

\textbf{Output quality} \( M_{\text{quality}}(j) \)  
quantifies task accuracy in a normalized range and is relevant for perception, detection, or estimation tasks. Compute tiers with degraded precision, reduced memory, or radiation susceptibility may show lower quality.

\textbf{Silent data corruption probability} (optional) is relevant in radiation-abundant environments such as GEO or interplanetary space, where increased particle flux can cause undetected bit-flips, potentially compromising result integrity.

\textbf{Compute availability} \( M_{\text{comp\_avail}}(j) \)  
measures the fraction of time a compute resource is operational and not undergoing downtime, maintenance, thermal throttling, or radiation resets. Higher availability indicates more predictable and reliable computing.

\subsubsection{Energy and Power Metrics}

\textbf{Energy per task} \( M_{\text{energy}}(j) \)  
is a critical constraint for power-constrained spacecraft. Onboard computing incurs a direct energetic cost, and high energy demands may reduce mission lifetime or interfere with other subsystems.

\textbf{Peak power draw} \( M_{\text{peak\_power}}(j) \)  
captures instantaneous power requirements that must be supported by the spacecraft power conditioning system. Excessive peak demand may exceed power limits even when average energy remains low.

\textbf{Power margin} \( M_{\text{power\_margin}}(j) \)  
quantifies available electrical headroom. Positive margins favor local computation, while negative margins make it infeasible.

\textbf{Generation capacity} \( M_{\text{power\_gen}}(j) \)  
describes the long-term power budget available for computational workloads and varies with orbital geometry and vehicle size.

\subsubsection{Thermal Metrics}

\textbf{Thermal margin} \( M_{\text{thermal}}(j) \)  
represents the temperature headroom before exceeding hardware thermal limits. Onboard and orbital systems often operate with tight margins due to limited radiative area, making compute-intensive tasks potentially unsafe during periods of high solar load or when thermal control is constrained.

\subsubsection{Communication Metrics}

\textbf{Link availability} \( M_{\text{link}}(j) \)  
measures the probability that bandwidth and link quality meet operational thresholds. Compute tiers requiring substantial uplink/downlink traffic (e.g., ground processing) depend heavily on link reliability.

\textbf{Contact duty cycle} \( M_{\text{contact}}(j) \)  
encodes the fraction of mission time during which communication is possible. Space-to-ground links may be intermittent, making remote computation infeasible for continuous or urgent tasks.

\textbf{Data reduction ratio} \( M_{\text{reduction}}(j) \)  
describes how much computation reduces data volume. Tasks with high reduction ratios benefit strongly from onboard computation because transmitting raw data is expensive or impossible under bandwidth constraints.

\textbf{Throughput capacity} (optional) captures effective data rates for communication-intensive tasks.

\subsubsection{Cost and Operational Metrics}

\textbf{Marginal cost per task} \( M_{\text{cost}}(j) \)  
represents direct or amortized financial cost associated with performing a computation at tier \( j \). Cloud resources tend to be inexpensive, while onboard compute has high marginal opportunity cost due to limited energy, mass, and power budgets.

\textbf{Operational overhead} \( M_{\text{ops}}(j) \)  
measures the human workload needed to maintain or supervise computing resources. Higher ops burden reduces desirability for sustained mission operations, particularly for autonomous spacecraft.

\subsubsection{Platform and Design Metrics}

\textbf{Orbital altitude} \( M_{\text{alt}}(j) \)  
affects propagation delay, radiation exposure, and contact opportunities. Although not inherently good or bad, altitude is an underlying driver for other metrics and can be considered directly as a criterion.

\textbf{Compute subsystem mass} \( M_{\text{mass}}(j) \)  
reflects the mass penalty of hosting additional onboard compute hardware. Lower mass supports more efficient mission architectures and reduces launch cost.

\textbf{Compute throughput capacity} \( M_{\text{throughput}}(j) \)  
indicates the rate at which a compute tier can process tasks and determines whether it can support the mission’s computational load.

\subsubsection{Regulatory Metric}

\textbf{Regulatory feasibility} \( R_{\text{ok}}(j) \)  
is a binary indicator capturing legal, privacy, or mission-specific restrictions on processing certain data at a given compute tier. This criterion enforces a hard constraint that supersedes utility considerations.

\subsubsection{Summary of Metric Rationale}

Collectively, these variables represent the principal physical, operational, informational, and economic factors governing the feasibility and desirability of performing compute in space or on the ground. Each metric contributes to a scalar utility that summarizes the trade space while preserving interpretability and mission relevance. The optimization function integrates these factors into a unified decision-making framework capable of handling incomplete data and heterogeneous compute environments.

\subsection{Metric Normalization}

Metrics differ in physical units and ranges. To compare them on a common scale, each metric \( M_i(j) \) is normalized to a dimensionless score \( s_i(j) \in [0,1] \). For criteria where higher values are preferred:
\begin{equation}
s_i(j) 
= \frac{M_i(j) - m_i^{\min}}{m_i^{\max} - m_i^{\min}}.
\label{eq:normalize_high}
\end{equation}

For criteria where lower values are preferred:
\begin{equation}
s_i(j) 
= \frac{m_i^{\max} - M_i(j)}{m_i^{\max} - m_i^{\min}}.
\label{eq:normalize_low}
\end{equation}

Scores are clipped to the interval \([0,1]\):
\[
s_i(j) = \min\{1,\max\{0,s_i(j)\}\}.
\]

\subsection{Utility Computation from Known Metrics}

Not all metrics are available for all tiers. Let \( K_j \) denote the subset of metrics known for tier \( j \). The \textit{base utility} is computed as a weighted average over known metrics:
\begin{equation}
U_{\text{base}}(j)
= \frac{\sum_{i \in K_j} w_i \, s_i(j)}
       {\sum_{i \in K_j} w_i}.
\label{eq:utility_base}
\end{equation}

\subsection{Robustness to Missing Data}

To avoid overestimating tiers with sparse information, we penalize tiers according to the fraction of known metric weight:
\begin{equation}
\phi(j) = \frac{W_{\text{known}}(j)}{W_{\text{total}}},
\end{equation}
where
\[
W_{\text{known}}(j) = \sum_{i \in K_j} w_i, \qquad
W_{\text{total}} = \sum_{i \in I} w_i.
\]

The \textit{effective utility} used for decision making is:
\begin{equation}
U_{\text{eff}}(j)
= U_{\text{base}}(j)
  - \lambda \left( 1 - \phi(j) \right),
\label{eq:utility_eff}
\end{equation}
where \( \lambda \in [0,1] \) is a configurable penalty parameter reflecting risk tolerance.

\subsection{Feasibility Constraints}

Mission requirements impose hard feasibility conditions. Tier \( j \) is infeasible if any of the following hold:
\begin{align}
M_{\text{latency}}(j) &> D_{\text{req}}, \\
M_{\text{success}}(j) &< R_{\text{req}}, \\
M_{\text{quality}}(j) &< Q_{\text{req}}, \\
M_{\text{cost}}(j)     &> C_{\text{req}}, \\
R_{\text{ok}}(j) &= 0.
\end{align}
In this case:
\[
U_{\text{eff}}(j) = -\infty.
\]


\subsection{Optimization Problem}

The compute-location selection is formulated as a single-step maximization over the feasible tiers. Each tier \( j \) receives an effective utility value \( U_{\text{eff}}(j) \), which incorporates normalized performance metrics, missing-data penalties, and feasibility constraints. The optimal compute tier is selected as
\begin{equation}
j^\star = \argmax_{j} \, U_{\text{eff}}(j).
\label{eq:argmax_choice}
\end{equation}

This scalarized multi objective formulation is computationally lightweight, interpretable, and directly extensible as additional metrics or constraints become relevant. It supports mission scenarios with incomplete metric observations and heterogeneous compute resources, while preserving mathematical transparency and operational practicality.

\section{Experimentation}
\label{sec:experiment}

We evaluate the presented optimization framework on two notional workloads. The first is an intrusion detection system (IDS) that must decide where to execute anomaly analytics based on data derived from the spacecraft's command and data handling (CDHS) architecture. The second is an abstraction of Google's Suncatcher program, in which a neural network training workload is placed in either a LEO TPU constellation, terrestrial data centers, or a hybrid split configuration.

\subsection{Intrusion Detection System Example}
\label{sec:ids_experiment}
For the IDS task, we provide the following metrics: p99 latency, success probability, output quality, energy per task, cost per task, link availability, operational burden, and data reduction ratio. Table~\ref{tab:ids_metrics} lists the numerical values used for each tier. Metrics defined in the complete optimization model but not available for this workload (peak power, power margin, thermal margin, contact duty cycle, compute availability, orbital altitude, and compute subsystem mass) are omitted. These metrics do not contribute to the utility computation for this experiment, and their absence only reduces the information fraction \(\phi(j)\), which triggers the penalty term in Eq.~\eqref{eq:utility_eff} without changing feasibility.

\begin{table}[h!]
\centering
\caption{IDS workload metrics per tier}
\label{tab:ids_metrics}
\scriptsize
\begin{tabular}{lcccc}
\hline
Metric & FC & ODC & GSE & TDC \\ \hline
Latency p99 (ms) & 90 & 180 & 600 & 320 \\
Success probability & 0.96 & 0.93 & 0.98 & 0.90 \\
Quality (0--1) & 0.93 & 0.91 & 0.95 & 0.88 \\
Energy per task (J) & 150 & 90 & 40 & 25 \\
Cost per task (USD) & 2.4 & 1.6 & 0.9 & 0.4 \\
Link availability & 0.78 & 0.88 & 0.97 & 0.99 \\
Ops minutes / 1000 tasks & 40 & 25 & 15 & 12 \\
Reduction ratio & 12 & 10 & 6 & 14 \\ \hline
\end{tabular}
\end{table}
\normalsize

The IDS scenario imposes a \SI{250}{\milli\second} p99 latency requirement, minimum success probability of 0.92, minimum output quality of 0.9, and a maximum marginal cost of \SI{3}{\$} per processed event. These thresholds are applied as hard feasibility constraints prior to utility computation. Any tier exceeding the latency or cost limits, or falling below the quality or reliability thresholds, is marked infeasible and assigned \(U_{\text{eff}} = -\infty\).

The metric weights for this experiment emphasize responsiveness and correctness: \(w_{\text{latency}} = 0.2\), \(w_{\text{success}} = 0.2\), \(w_{\text{quality}} = 0.15\), \(w_{\text{energy}} = 0.1\), \(w_{\text{cost}} = 0.1\), \(w_{\text{link}} = 0.15\), \(w_{\text{ops}} = 0.05\), and \(w_{\text{reduction}} = 0.05\). After normalizing the metrics and applying these weights, we compute the base utility \(U_{\text{base}}\), apply the missing-data penalty, and finally obtain the effective utility \(U_{\text{eff}}\) for each feasible tier.


\subsection{Suncatcher AI Training Workload}
\label{sec:suncatcher_experiment}

To illustrate a complementary trade space that stresses power, launch mass, and long running reliability, we formulate an example based on Google Suncatcher in which large scale neural network training can occur in LEO or on the ground. The candidate tiers are a dawn to dusk sun synchronous orbit TPU constellation (LEO\_TPU\_CLUSTER), a terrestrial TPU data center (GROUND\_TPU\_DC), a terrestrial GPU or CPU data center (GROUND\_GPU\_DC), and a hybrid split pipeline (HYBRID\_SPLIT) that divides training between orbit and ground.

For this workload we exercise the full metric set, including p99 latency, success probability, quality score, energy per training unit, peak power, power margin, thermal margin, link availability, contact duty cycle, data reduction ratio, cost per training unit, operational burden, compute availability, orbital altitude, and compute mass. Task level limits enforce \(D_{\text{req}} = \SI{120}{\milli\second}\), \(R_{\text{req}} = 0.993\), \(Q_{\text{req}} = 0.985\), and \(C_{\text{req}} = \$14\) per training unit. The raw metric values are summarized in Table~\ref{tab:suncatcher_metrics}.

\begin{table}[h!]
\centering
\caption{Suncatcher AI training metrics per tier. \textit{Metrics are hypothetical and should not be conflated with a genuine Google Suncatcher system.}}
\label{tab:suncatcher_metrics}
\scriptsize
\begin{tabular}{lcccc}
\hline
Metric & LEO\_TPU\_CLUSTER & GROUND\_TPU\_DC & GROUND\_GPU\_DC & HYBRID\_SPLIT \\ \hline
Latency p99 (ms)           & 80      & 40      & 60      & 100 \\
Success probability        & 0.995   & 0.999   & 0.998   & 0.993 \\
Quality score              & 0.990   & 0.990   & 0.985   & 0.990 \\
Energy per unit (J)        & \(1.10\times10^6\) & \(1.00\times10^6\) & \(1.50\times10^6\) & \(1.20\times10^6\) \\
Peak power (W)             & 250{,}000 & 300{,}000 & 350{,}000 & 270{,}000 \\
Power margin (W)           & 150{,}000 & 80{,}000  & 60{,}000  & 100{,}000 \\
Thermal margin (\si{\celsius}) & 15  & 10      & 8       & 12 \\
Link availability          & 0.980   & 0.999   & 0.997   & 0.960 \\
Contact duty cycle         & 0.70    & 1.00    & 1.00    & 0.85 \\
Reduction ratio            & 20      & 1       & 1       & 10 \\
Cost per unit (USD)        & 10      & 12      & 15      & 13 \\
Ops min.\,/ 1000 units     & 40      & 20      & 25      & 50 \\
Compute availability       & 0.985   & 0.999   & 0.998   & 0.970 \\
Orbital altitude (km)      & 550     & 0       & 0       & 550 \\
Compute mass (kg)          & 575     & 2{,}000 & 2{,}200 & 1{,}500 \\ \hline
\end{tabular}
\normalsize
\end{table}

For this case study the weights emphasize latency, reliability, and operational considerations: \(w_{\text{latency}} = 0.10\), \(w_{\text{success}} = 0.15\), \(w_{\text{quality}} = 0.10\), \(w_{\text{energy}} = 0.10\), \(w_{\text{peak}} = 0.05\), \(w_{\text{margin}} = 0.05\), \(w_{\text{thermal}} = 0.05\), \(w_{\text{link}} = 0.08\), \(w_{\text{duty}} = 0.04\), \(w_{\text{reduction}} = 0.06\), \(w_{\text{cost}} = 0.12\), \(w_{\text{ops}} = 0.03\), \(w_{\text{availability}} = 0.04\), \(w_{\text{altitude}} = 0.02\), and \(w_{\text{mass}} = 0.01\). As in the IDS experiment, metrics are normalized, aggregated into \(U_{\text{base}}\), and adjusted by the missing data penalty to obtain \(U_{\text{eff}}\), subject to the task level feasibility constraints.

\section{Results}
\label{sec:results}

\subsection{IDS Results}
\label{sec:ids_results}

The utilities produced by the optimization for the IDS workload are shown in Table~\ref{tab:ids_results}. The two tiers that satisfy all task level requirements are the FC and the ODC. Their effective utilities differ by only 0.003, with ODC achieving the highest score. Although FC offers the lowest latency, ODC benefits from lower energy cost, lower operational burden, and higher link availability. Under the chosen weights, these advantages slightly outweigh FC's latency benefit. Both feasible tiers incur a small missing data penalty due to omitted metrics.

GSE and TDC violate the \SI{250}{\milli\second} latency constraint (\SI{600}{\milli\second} and \SI{320}{\milli\second}, respectively), and are therefore rendered infeasible. As feasibility is evaluated before utility scoring, these tiers are assigned \(U_{\text{eff}} = -\infty\) and excluded from comparison.

\begin{figure}
    \centering
    \includegraphics[width=0.5\linewidth]{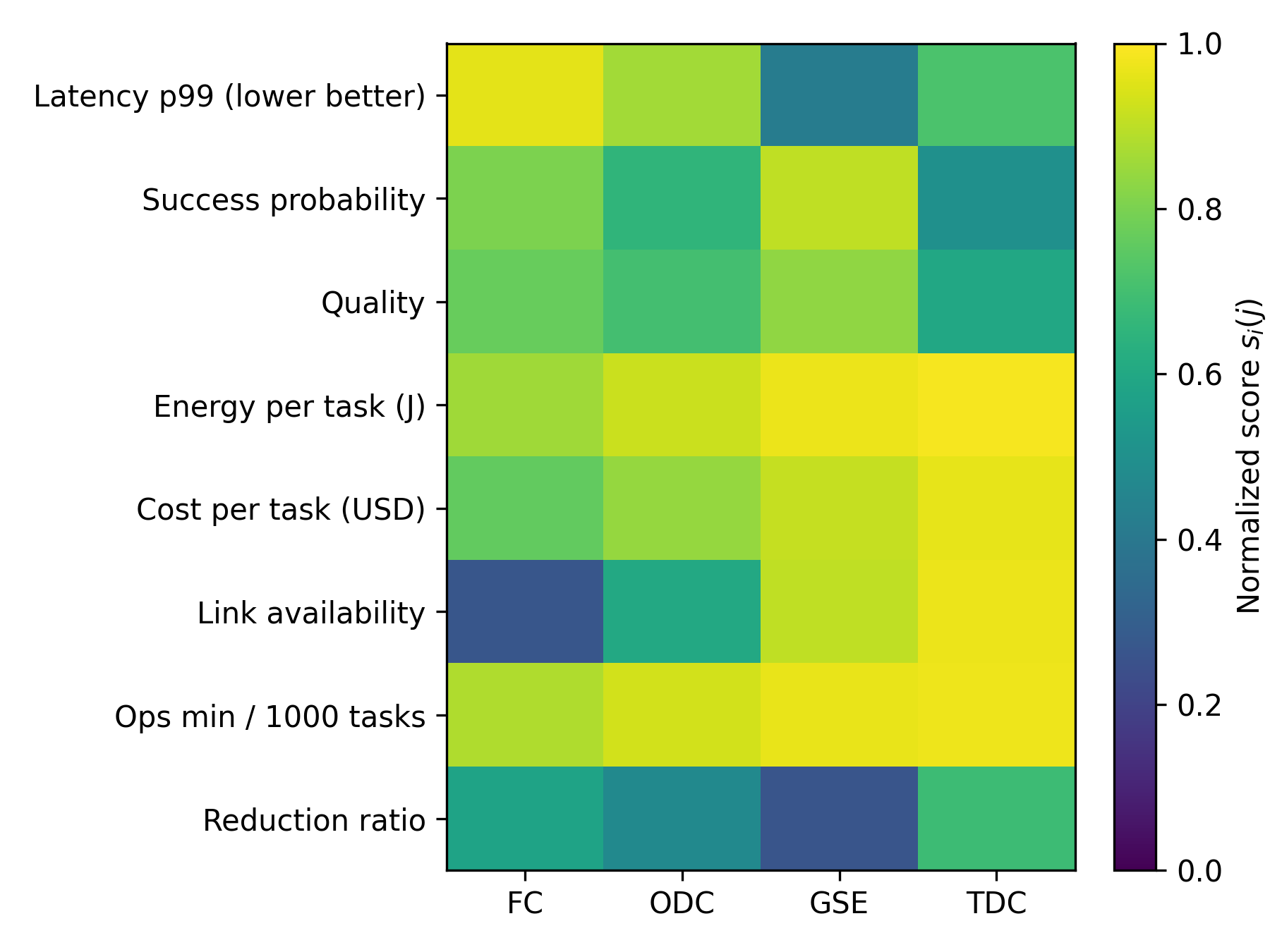}
    \caption{Normalized per-metric scores for the IDS workload.}
    \label{fig:IDS_heatmap}
\end{figure}

\begin{figure}[h!]
    \centering
    \includegraphics[width=0.8\linewidth]{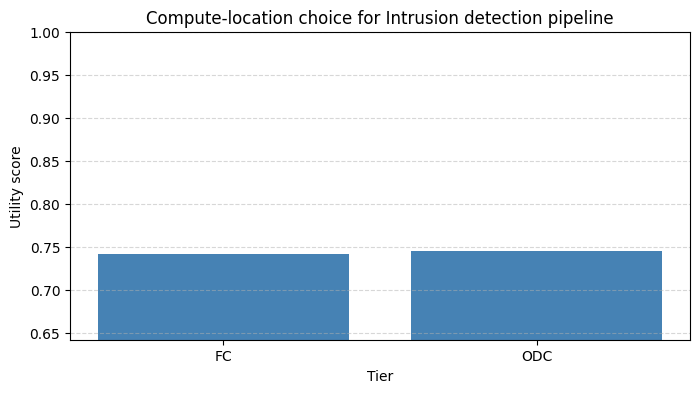}
    \caption{Penalized utility scores for the IDS workload across the flight computer (FC) and orbital data center (ODC). Ground station edge (GSE) and terrestrial data center (TDC) are omitted because their p99 latencies exceeded the 250\,ms requirement, yielding infeasible scores with \(U_{\text{eff}} = -\infty\).}
    \label{fig:tier_scores}
\end{figure}

Figure~\ref{fig:tier_scores} plots the effective utilities for the feasible tiers and normalized per-metric scores are depicted as a heatmap in \ref{fig:IDS_heatmap}. Together with Tables~\ref{tab:ids_metrics} and \ref{tab:ids_results}, this illustrates how the optimization identifies the orbital data center as the preferred compute tier for this IDS workload.

\begin{table}[h!]
\centering
\caption{Computed utilities for the IDS workload.}
\label{tab:ids_results}
\begin{tabular}{lccc}
\hline
Tier & Feasible? & $U_{\text{base}}$ & $U_{\text{eff}}$ \\ \hline
FC  & Yes & 0.806 & 0.742 \\ 
ODC & Yes & 0.801 & 0.745 \\ 
GSE & No (latency) & -- & $-\infty$ \\ 
TDC & No (latency) & -- & $-\infty$ \\ \hline
\end{tabular}
\end{table}

\subsection{Suncatcher AI Training Results}
\label{sec:suncatcher_results}

The resulting utilities for the Suncatcher workload are summarized in Table~\ref{tab:suncatcher_results}. Under the specified weights and constraints, the terrestrial TPU data center (GROUND\_TPU\_DC) attains the highest effective utility with \(U_{\text{eff}} = 0.784\). The LEO TPU constellation (LEO\_TPU\_CLUSTER) remains competitive with \(U_{\text{eff}} = 0.675\), supported by strong data reduction and power margins despite higher latency and operational burden. The hybrid split pipeline (HYBRID\_SPLIT) incurs coordination and operations penalties, yielding a lower effective utility of 0.352. The ground GPU tier (GROUND\_GPU\_DC) is infeasible because its cost per training unit exceeds the \$14 limit, so it is assigned \(U_{\text{eff}} = -\infty\).

\begin{table}[h!]
\centering
\caption{Computed utilities for the Suncatcher workload.}
\label{tab:suncatcher_results}
\begin{tabular}{lccc}
\hline
Tier              & Feasible? & Reason                            & $U_{\text{eff}}$ \\ \hline
LEO\_TPU\_CLUSTER & Yes       & --                                & 0.675 \\
GROUND\_TPU\_DC   & Yes       & --                                & 0.784 \\
GROUND\_GPU\_DC   & No        & Cost per unit \(> \SI{14}{\$}\)   & $-\infty$ \\
HYBRID\_SPLIT     & Yes       & --                                & 0.352 \\ \hline
\end{tabular}
\end{table}

\begin{figure}
    \centering
    \includegraphics[width=0.5\linewidth]{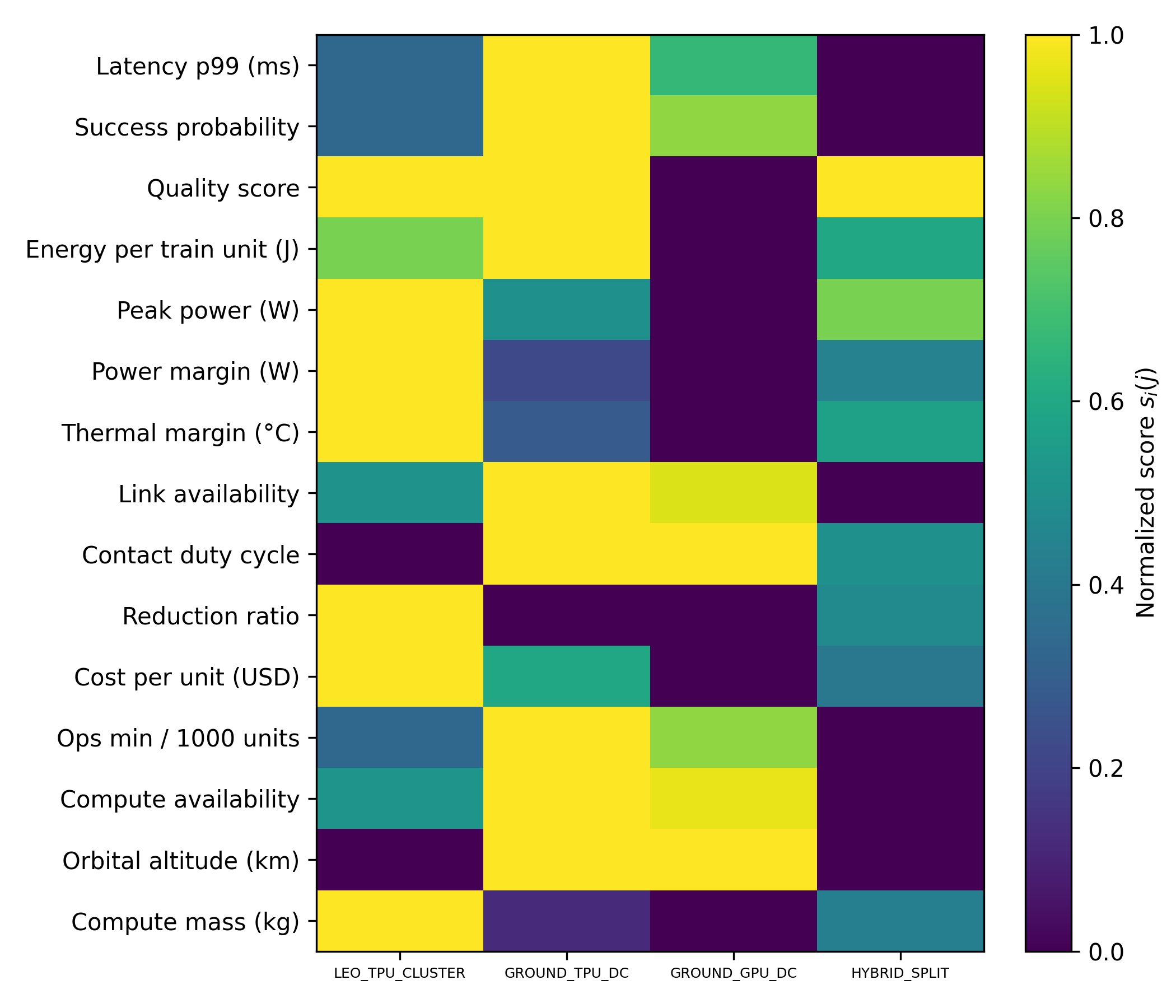}
    \caption{Normalized per-metric scores for the Suncatcher AI training tiers}
    \label{fig:suncatcherHeatmap}
\end{figure}

\begin{figure}[h!]
    \centering
    \includegraphics[width=0.75\linewidth]{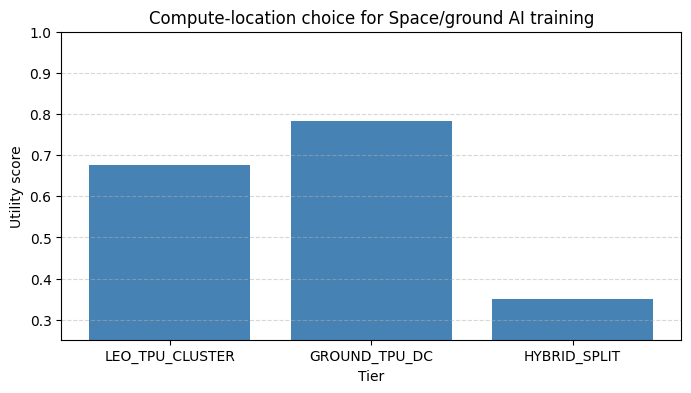}
    \caption{Effective utility scores for the Suncatcher AI training workload. GROUND\_GPU\_DC is omitted because its cost exceeds the \SI{14}{\$} per unit constraint, yielding \(U_{\text{eff}} = -\infty\).}
    \label{fig:suncatcher_scores}
\end{figure}

Figure~\ref{fig:suncatcher_scores} and Figure \ref{fig:suncatcherHeatmap} visualizes the effective utilities for the feasible Suncatcher tiers. Together with Table~\ref{tab:suncatcher_metrics}, this example shows how the same optimization framework can compare orbital and terrestrial training deployments, accounting for power, mass, cost, and communication structure in addition to traditional performance metrics.

\section{Discussion}
\label{sec:discussion}

This paper was intended as a quantitative treatment of a question that is often handled informally in mission design, namely, where a given workload should execute along the space to ground compute continuum. The optimization model makes that question explicit by forcing all relevant metrics into a shared numerical representation and by requiring that trade offs among latency, power, link characteristics, cost, and platform properties be stated as weights and constraints rather than left implicit.

Mission engineers do not typically frame this choice as an optimization problem. Decisions about onboard versus ground processing are often driven by a combination of rules of thumb, heritage, and programmatic or security constraints that are evaluated qualitatively. The framework in this paper does not replace that practice, it provides a structured way to expose which assumptions are driving a particular choice and how sensitive the decision is to changes in requirements or architecture.

Several constraints that practitioners apply in real programs appear only indirectly in the current model. Security is a prominent example. A design may be eliminated because cryptographic key management, key reload, or hardware security module support is not available at a particular tier, even if all performance metrics are favorable. In such a case, the relevant constraint is not a small adjustment to utility, it is an absolute prohibition on processing specific data in that location. Similar effects arise from export control rules, data sovereignty policies, or organizational boundaries between different mission stakeholders. These show up today as qualitative gates in design reviews rather than as explicit mathematical predicates.

Cost is another dimension where current practice and the model intersect only partially. The framework includes marginal cost per task, which is useful for comparing recurring processing expenses across tiers. Mission cost decisions often couple that quantity with non recurring engineering cost, launch cost, and life cycle support arrangements that are negotiated at a program level. Engineers also choose where to compute at specific points in the design timeline, for example at proposal, preliminary design review, or critical design review, when technology maturity and vendor offerings differ from what will be available at launch. The static formulation presented here does not yet reflect those temporal and organizational constraints, although it can be used as a diagnostic tool to understand how much of a decision is driven by measurable performance and how much by external factors.

Overall, the results from the IDS and Suncatcher examples demonstrate that a relatively simple scalarized formulation can reproduce many of the intuitive trade offs that mission teams currently make, while also making it clear where qualitative constraints would overrule an otherwise attractive placement.

\section{Future Work}
\label{sec:future_work}

Future work falls into three main directions that connect this quantitative framework to real missions and existing engineering practice.

First, the model should be exercised on concrete science, commercial, and defense mission concepts in collaboration with flight and ground segment teams. The IDS and Suncatcher case studies provide a proof of concept, but they do not reflect the full complexity of operational missions. Applying the framework to candidate missions, varying orbital parameters, payload types, and link architectures, and then comparing the recommended compute locations with current baselines would help determine when orbital data centers, onboard processing, or terrestrial infrastructure are genuinely optimal.

Second, the decision process itself should be studied and encoded. The present work treats compute placement as a clean optimization problem with well defined metrics, weights, and constraints. Mission engineers often make these decisions through a sequence of design trades, subsystem reviews, and programmatic negotiations that are not captured in a single objective function. Mapping how teams currently decide where to place compute, including which constraints are treated as hard qualitative gates, would inform how regulatory, security, and organizational factors should be incorporated. Cryptographic keying infrastructure provides a concrete example. If a mission cannot support re keying or key distribution at a given tier, then certain workloads cannot legally or operationally run there, regardless of utility. Future versions of the model should represent such constraints explicitly rather than folding them into generic regulatory feasibility terms.

Third, the optimization framework will be extended to support explicit multi objective analysis rather than relying solely on the current scalarized utility formulation. In its present form, the model collapses all metrics into a single weighted score and selects the tier that maximizes that value, which is appropriate for decision automation but does not expose the full structure of the trade space. Introducing Pareto front analysis would allow mission teams to visualize competing objectives directly, identify nondominated compute tiers, and understand cases where a tier with a lower aggregated utility may still be preferable for reasons such as autonomy, reduced operational dependency, or architectural simplicity. For example, the orbital data center may maximize the scalar utility, while the flight computer remains the nondominated choice when minimizing external reliance or ensuring deterministic availability. Adding Pareto front generation would therefore complement the existing formulation by enabling engineers to examine multi metric tradeoffs without committing to any specific weighting scheme.

\section{Conclusion}
\label{sec:conclusion}

This paper has proposed a quantitative framework for deciding when to compute in space and when to rely on ground or orbital infrastructure. The model enumerates a set of empirically measurable metrics, normalizes them, combines them through a transparent utility function with explicit handling of missing data, and enforces feasibility through hard constraints on latency, reliability, quality, cost, and regulatory status. Two case studies, an intrusion detection workload and a Suncatcher style training workload, show how the same structure can support very different tasks and architectures while preserving interpretability.

The formulation does not capture every constraint that mission engineers apply in practice, particularly qualitative security, programmatic, and organizational considerations. It also does not model when in the design process compute placement decisions are made or how technology evolves over the mission life cycle. Despite these limitations, the framework provides a starting point for bringing rigor and repeatability to a class of decisions that will become more frequent as orbital data centers and advanced onboard processors mature. Continued work with real missions, and closer integration with existing engineering workflows, will be required to determine how this kind of optimization should sit alongside the qualitative judgments that ultimately shape space system architectures.

\bibliography{sample}

@misc{fischbacher2025towards,
  title = {Towards a future space-based, highly scalable AI infrastructure system design},
  author = {Fischbacher, Thomas},
  year = {2025},
  month = {nov},
  eprint = {2511.19468},
  archivePrefix = {arXiv},
  primaryClass = {cs.DC},
  doi = {10.48550/arXiv.2511.19468},
  url = {https://arxiv.org/abs/2511.19468},
}

@misc{peraspera2025realities,
  title = {Realities of Space-Based Compute},
  author = {Goldin, Duffy},
  year = {2025},
  month = {may},
  day = {12},
  organization = {Per Aspera},
  url = {https://www.peraspera.us/realities-of-space-based-compute/},
}

@article{liu2025computing,
  title = {Computing over Space: Status, Challenges, and Opportunities},
  author = {Liu, Yaoqi and Han, Yinhe and Li, Hongxin and Gu, Shuhao and Qiu, Jibing and Li, Ting},
  journal = {Engineering},
  volume = {54},
  number = {Suppl. C},
  pages = {20--25},
  year = {2025},
  month = {nov},
  doi = {10.1016/j.eng.2025.00299-1},
  url = {https://www.sciencedirect.com/science/article/pii/S2095809925002991},
}

@misc{feilden2024why,
  title = {Why we should train AI in space},
  author = {Feilden, Ezra and Oltean, Adi and Johnston, Philip},
  year = {2024},
  month = {sep},
  organization = {Starcloud},
  howpublished = {\url{https://starcloudinc.github.io/wp.pdf}},
  note = {White paper v1.03. Originally published by Lumen Orbit},
  url = {https://starcloudinc.github.io/wp.pdf},
}

@misc{axiomspace2025inspace,
  title = {In-Space Data \& Security},
  year = {2025},
  organization = {Axiom Space},
  url = {https://www.axiomspace.com/in-space-data-security},
}

@misc{callison2025beyond,
  title = {Beyond the horizon: cost-driven strategies for space-based data centers},
  author = {Callison, John David and Minafra, Joseph},
  year = {2025},
  month = {dec},
  day = {8},
  publisher = {SpaceNews},
  url = {https://spacenews.com/beyond-the-horizon-cost-driven-strategies-for-space-based-data-centers/},
}

@inproceedings{ThummalaFalco2026WhenToComputeInSpace,
  author    = {Thummala, Rajiv and Falco, Gregory},
  title     = {When to Compute in Space},
  booktitle = {AIAA SciTech Forum},
  year      = {2026},
  address   = {Orlando, FL, USA},
  month     = jan,
}

\end{document}